\documentclass[prd,preprint,superscriptaddress]{revtex4}
\usepackage{amsmath}
\usepackage{amssymb}
\usepackage{revsymb}
\usepackage{graphicx}
\newcommand{\be}{\begin{equation}}
\newcommand{\ee}{\end{equation}}
\newcommand{\ben}{\begin{eqnarray}}
\newcommand{\een}{\end{eqnarray}}

\newcommand{\gsim}[2]{
\setlength{\unitlength}{12pt}
\begin{picture}(1.4,1.)
\put(.7,-0.3){\makebox(0.0,1.)[t]{$>$}}
\put(.7,-0.3){\makebox(0.0,1.)[b]{$\sim$}}
\end{picture}#2}

\begin{document}
\title{Big rip avoidance via black holes production}
\author{J\'{u}lio C. Fabris\footnote{E-mail address:
fabris@pq.cnpq.br.}} \affiliation{Departamento de F\'{\i}sica,
Universidade Federal do Esp\'{\i}rito Santo, CEP 29060-900
Vit\'{o}ria, Esp\'{\i}rito Santo, Brasil}
\author{Diego Pav\'{o}n\footnote{E-mail address: diego.pavon@uab.es}}
\affiliation{Departamento de F\'{\i}sica, Universidad Aut\'{o}noma
de Barcelona, 08193 Bellaterra (Barcelona), Spain}

\begin{abstract}
We consider a cosmological scenario in which the expansion of the
Universe is dominated by phantom dark energy and black holes which
condense out of the latter component. The mass of  black holes
decreases via Hawking evaporation and by accretion of phantom
fluid but new black holes arise continuously whence the overall
evolution can be rather complex. We study the corresponding
dynamical system to unravel this evolution and single out
scenarios where the big rip singularity does not occur.
\end{abstract}

\maketitle

\section{Introduction}
Phantom dark energy fields are characterized by violating the
dominant energy condition, $\rho + p > 0$. Thereby the
conservation equation, $\dot{\rho} + 3 H (\rho + p) = 0$, has the
striking consequence that the energy density increases with
expansion \cite{caldwell1,caldwell2}. In the simplest case of a
constant ratio $w \equiv p/\rho$ one has $\rho \propto
a^{-3(1+w)}$, where $1+w < 0$, while the scale factor obeys $a(t)
\propto (t_{*} - t)^{-n}$ with $n = - 2/[3(1+w)]$ and $t \leq
t_{*}$, being $t_{*}$ the ``big rip" time (at which the scale
factor diverges). However, as we shall show below, the big rip may
be avoided if black holes are produced out of the phantom fluid at
a sufficiently high rate. One may think of three different
mechanisms by which phantom energy goes to produce black holes.

$(i)$ In phantom dominated universes more and more energy is
continuously pumped into any arbitrary spatial three-volume of
size $R$. Hence, the latter will increase as $a$ while the mass,
$M$, inside it will go up as as $a^{m+3}$ -with $m = 3\,|1+w|$.
Therefore, the ratio $M/R$ will augment with expansion whenever $m
> -2$ (i.e., $w < -1/3$, dark energy in general). As a consequence
of the energy being pumped faster than the volume can expand, the
latter will eventually contain enough energy to become a black
hole. This is bound to occur as soon as $ \rho \, R^{3} \geq R/(2
G)$ (i.e., when $R \geq (2G\, \rho)^{-1/2}$).

$(ii)$ As statistical mechanics tells us, equilibrium thermal
fluctuations obey $<(\delta E)^{2}> = k_{B} \, C_{V} \, T^{2}$,
where $\delta E = E - <E>$ is the energy fluctuation at any given
point of the system around the mean value $<E>$, $C_{V}$ the
system heat capacity at constant volume, $k_{B}$ Boltzmann's
constant, and $< ... >$ denotes ensemble average \cite{textbook}.
As demonstrated in \cite{db}, unlike normal matter,  dark energy
gets hotter with expansion according to the law $T \propto
a^{-3w}$. The latter follows from integrating the temperature
evolution equation $\dot{T}/T = -3H (\partial p/\partial
\rho)_{n}$ which, on its turn, can be derived from Gibbs' equation
$T \, dS = d(\rho/n) + p \, d(1/n)$ and the condition that the
entropy be a state function, i.e., $\partial^{2}S/(\partial T \,
\partial n) = \partial^{2}S/(\partial n \, \partial T)$
-see \cite{db} for details. Therefore, it is natural to expect
that the phantom fluid will be very hot at the time when it starts
to dominate the expansion. Since the phantom temperature grows
unbounded so it will the energy fluctuations as well -a
straightforward calculation yields $<(\delta E)^{2}> \propto
a^{-3(1+2w)}$. Eventually they will be big enough to collapse and
fall within their Schwarzschild radius. (This simple mechanism was
applied to hot thermal radiation well before phantom energy was
introduced \cite{piran-wald}).

$(iii)$ As demonstrated by Gross {\it et al.}
\cite{quantumtunnelling}  thermal radiation can give rise to a
copious  production of black holes of mass $ T^{-1}$ whenever $T$
is sufficiently high. In this case, black holes nucleate because
the small perturbations around the Schwarzschild instanton involve
a negative mode. This produces an imaginary part of the free
energy that can be interpreted as an instability of the hot
radiation to quantum tunnel into black holes. A simpler (but less
rigorous) treatment can be found in Ref. \cite{Kapusta}. One may
speculate that black holes might also come into existence by
quantum tunnelling of hot phantom fluid similarly as hot radiation
did in the very early Universe.

Notice that mechanisms $(i)$ and $(iii)$ differ from the
conventional gravitational collapse at zero temperature -the
latter does not produce black holes when the fluid has $w < -1/3$,
see e.g. \cite{cai-wang}.

Obviously, one may wonder whether if phantom can produce particles
other than black holes. In actual fact, there is no reason why
dark energy in general (not just phantom) should not be coupled to
other forms of matter. However, its coupling to baryonic matter is
highly constrained by measurements of local gravity
\cite{localgravity} but not so to dark matter. In any case, we
will not consider such possibility because it would introduce an
additional variable in our system of equations below (Eqs.
(\ref{bhconsvbis})- (\ref{dotm1})) and would increase greatly the
complexity of the analysis.

The number black holes in the first generation (assuming that all
those initially formed arise simultaneously) will be $N_{i} =
(a/R)_{i}^{3} = C\,  a ^{3-9(1+w)/2}$ with $C = [2 G \rho_{0} \,
a_{0}^{3(1+w)}]^{3/2}$ -here the subscript zero signals the
instant at which phantom dark energy starts to overwhelmingly
dominate all other forms of energy-, and we may, reasonably,
expect that they will constitute a pressureless fluid. Later on,
more black holes may be formed but if much of the phantom energy
has gone into black holes, then the second generation will not
come instantly (as there will be less available phantom energy
than at $t = t_{i}$).

Further, the black holes will accrete phantom energy and lose mass
at a rate $\dot {M} = -16 \pi M^{2} \, \dot{\phi} ^{2}$ regardless
of the phantom potential, $V(\phi)$ \cite{babichev}. (Notice that
for scalar phantom fields $\rho + p = - \dot{\phi} ^{2}$).
Additional mass will be lost to Hawking radiation
\cite{hradiation}.

The system of equations governing this complex scenario is
\\
\be \dot{\rho}_{bh} + 3H \rho_{bh} = \Gamma \, \rho_{x} + 16 \pi
\, n M^{2} \, (1+w)  \rho_{x} \, - n\, \frac{\alpha}{M^{2}}\, ,
\label{bhconsv}
 \ee
 \be
\dot{\rho}_{x} + 3(1+w)H\rho_{x} = - \Gamma \, \rho_{x} - 16 \pi
\, n M^{2} \,  (1+w) \rho_{x} \, ,
 \label{phconsv}
 \ee
 \be
 \dot{\rho}_{\gamma} + 4 H \rho_{\gamma} = n
 \frac{\alpha}{M^{2}}\, ,
\label{gammaconsv}
 \ee
\be 3 H^{2} = \rho_{x} + \rho_{bh}+ \rho_{\gamma} \, ,
\label{friedmann} \ee \be \dot{M} = 16 \pi \, M^{2} (1+w) \rho_{x}
- \frac{\alpha}{M^{2}}
\label{dotm} \ee
\\
(subscripts $bh$, $x$, and $\gamma$ stand for black hole, phantom,
and radiation components, respectively, and we use units in which
$8\pi G = c = \hbar = 1$). Here, $\Gamma =$ constant $>0$ denotes
the rate of black hole formation, $n = N/a^{3}$ the number density
of black holes, $\alpha$ a positive constant, and $H = \dot{a}/a$
the Hubble function. The first term in (\ref{dotm}) corresponds to
the mass loss rate of a single black hole via phantom accretion;
the second term to spontaneous Hawking radiation \cite{donpage}.
For simplicity, we assume that the black holes solely emit
relativistic particles (i.e., a fluid with equation of state
$p_{\gamma} = \rho_{\gamma}/3$). This explains Eq.
(\ref{gammaconsv}). The second term on the right of Eq.
(\ref{phconsv})  ensures the energy conservation of the overall
phantom plus black hole fluid in the process of phantom accretion
-the mass loss of black holes must go into phantom energy.

At first glance, depending on the values assumed by the different
parameters, $\Gamma$, $w$, and $\alpha$, two very different
outcomes seem possible: $(i)$ a big rip singularity, if the
phantom energy eventually gets the upper hand; and $(ii)$ a
quasi-equilibrium situation between the black hole and phantom
fluids, if none of them comes to dominate the expansion -in this
case the  big rip cannot be guaranteed since the overall equation
of state will not be constant.

There are five equations and, seemingly, six unknowns, namely,
($\rho_{bh}$, $\rho_{x}$, $\rho_{\gamma}$, $H$, $n$ and $M$) but,
in actual fact only five ($\rho_{bh}$, $\rho_{x}$,
$\rho_{\gamma}$, $a$, and $M$) as $n$ can be written as
$\rho_{bh}/M$.

To make things easier, we shall further assume that the black
holes are massive enough to safely neglect Hawking evaporation and
that, initially, there was no radiation present. The latter
assumption is not much unrealistic, the fast expansion redshifts
away radiation and dust matter very quickly. (The black-hole fluid
also has the equation of state of dust but, as said above, black
holes are continuously created out of phantom). Thus, we can
dispense with Eq. (\ref{gammaconsv}) and the last term on the
right hand side of Eqs. (\ref{bhconsv}) and (\ref{dotm}).

One may  wonder whether the black holes may coalesce leading to
bigger black holes and produce a big amount of relativistic
particles. We believe we can safely ignore this possibility since
the fast expansion renders the chances of black holes encounters
highly unlikely.

Accordingly, the system of equations reduces to
\\
 \be
 \dot{\rho}_{bh} + 3H \rho_{bh} = \Gamma \rho_{x} + 16 \pi \,
(1+w) M \, \rho_{bh} \, \rho_{x}  \, , \label{bhconsvbis}
 \ee
 \be
\dot{\rho}_{x} + 3(1+w)H \rho_{x} = - \Gamma \rho_{x} - 16 \pi \, M
(1+w) \rho_{bh} \, \rho_{x}  \, ,
 \label{phconsvbis}
 \ee
\be 3 H^{2} = \rho_{x} + \rho_{bh} \, , \label{friedmann1}
\ee
\be
\dot{M} = 16 \pi \, M^{2} (1+w)\rho_{x}\,  .
\label{dotm1}
\ee
\\
Thus,  we may choose the  three unknowns, $\rho_{bh}$, $\rho_{x}$
and $M$  -since $H$ is linked to $\rho_{bh}$ and $\rho_{x}$ by the
constraint Eq. (\ref{friedmann1}).

Let us assume a phantom dominated universe (i.e., no other energy
component enters the picture initially). Sooner or later a first
generation of black holes will arise; then, the question arises:
``will new black holes condense out of phantom (second generation)
before the first generation practically disappears eaten by the
phantom fluid and consequently, the Universe become forever
dominated by a mixture of phantom and black holes, or the black
holes will disappear before they can dispute the phantom the
energy dominance of cosmic expansion?"

To ascertain this we shall apply, in the next Section, the general
theory of dynamical systems  \cite{sansone} to the above set of
equations (\ref{bhconsvbis})--(\ref{dotm1}). We shall  analyze the
corresponding critical points at the finite region as well as at
infinity. As we will see, due to the existence of a critical point
in the finite region of the phase portrait (connected with the
creation rate, $\Gamma$, of new black holes), there exist
solutions that instead of ending up at the big rip tend
asymptotically to a Minkowski spacetime. Hence, in some cases,
depending on the initial conditions, the big rip singularity can
be avoided thanks to the formation of black holes out of the
phantom fluid.

Before going any further, it is fair to say that, in reality, no
one knows for certain if phantom fluids have a place in Nature:
they may suffer from quantum instabilities \cite{instability},
although certain phantom models based in low--energy  effective
string theory may avoid them \cite{piazza}. On the other hand,
observationally they are slightly more favored than otherwise
-though, admittedly, this support has dwindled away in the last
couple of years. In view of this unsettled situation, we believe
worthwhile to explore the main possible consequences its actual
existence may bring about on cosmic evolution.

\section{Dynamical study}
Here we apply the theory of dynamical systems to analyze the above
set of differential equations (\ref{bhconsvbis})--(\ref{dotm1}).
Using the total density $\rho_{T} = \rho_{bh} + \rho_{x}$, the
system can be recast as
\begin{eqnarray}
\label{e1}
\dot\rho_x &=& - 3(1 + w)\sqrt{\rho_T}\rho_x - \Gamma\rho_x - 16\pi(1 + w) M\rho_x(\rho_T - \rho_x) \, ,\\
\label{e2}
\dot\rho_T &=& -3(\rho_T^{3/2} + w\rho_T^{1/2}\rho_x) \, , \\
\label{e3} \dot M &=& 16\pi M^2(1 + w)\rho_x \, .
\end{eqnarray}
Its critical points follow from setting  $\dot\rho_x = \dot\rho_T
= \dot{M} = 0 \, $.

Three possible situations arise, namely:
\begin{enumerate}
\item $M = 0$ and $\rho_x = 0$; \item $M = 0$ and $\rho_x \neq 0$;
\item $M \neq 0$ and $\rho_x = 0$.
\end{enumerate}
In virtue of Eq. (\ref{e2}), the first one implies $\rho_T = 0$,
i.e., the origin. The second one corresponds to
\\
\begin{equation}
\rho_T = - w\rho_x \quad \mbox{and} \quad \sqrt{\rho_T} =-
\frac{\Gamma}{3(1 + w)} \, .
\end{equation}
Since both densities must be semi-positive definite, we have that
$w < -1$ which is consistent with the assumption of phantom fluid.
For fluids satisfying the dominant energy condition this critical
point does not exist. Finally, the third case also implies the
origin as, by virtue of Eq. (\ref{e2}), $\rho_T = 0$.

\subsection{The critical points at the finite region}
In the finite region, the obvious critical point  is the origin
$(M= \rho_{x} = \rho_{T} = 0)$, and, for the phantom case ($w
<-1$), the point given by
\begin{equation}
 \rho_x = - \frac{\Gamma^2}{9w(1 + w)^2} \, , \quad
\quad \rho_T = \frac{\Gamma^2}{9(1 + w)^2}\, , \quad \quad M = 0
\, .
\label{obvious}
\end{equation}

After linearizing the system of equations (\ref{e1})-(\ref{e3}),
we get
\\
\begin{eqnarray}
\delta\dot\rho_x &=& -\left[ \textstyle{3\over 2}(1 + w)\frac{\rho_x}{\sqrt{\rho_T}} + 16\pi M(1+w)\rho_x\right]\,
\delta\rho_T \nonumber\\
&-& \left[3(1+w)\sqrt{\rho_T} + \Gamma + 16\pi M(1 +w)(\rho_T - 2\rho_x)\right]\, \delta\rho_x \nonumber\\
&-& 16\pi(1 + w)(\rho_T - \rho_x)\rho_x \, \delta M \, , \label{system1}\\
\delta\dot\rho_T &=&  - \textstyle{3\over 2} \left[3\sqrt{\rho_T} + w\frac{\rho_x}{\sqrt{\rho_T}}\right]\,
\delta\rho_T - 3w \sqrt{\rho_T} \; \delta\rho_x \, , \label{system2}\\
\delta\dot M &=& 32\pi M(1 + w)\rho_x \, \delta M + 16\pi M^2(1
+w)\, \delta\rho_x \, \label{system3}.
\end{eqnarray}

\subsubsection{The critical point at the origin}
In this case, the system reduces to
\\
\be \delta\dot\rho_x = - \Gamma\delta\rho_x \, , \qquad
\delta\dot\rho_T = \delta\dot {M} = 0 \, . \ee Here we have made
the reasonable assumption that $\rho_x/\sqrt{\rho_T} = 0$  as
$\rho_x$, $\rho_T \rightarrow 0$. This critical point is an
attractor since its sole eigenvalue is negative.

\subsubsection{The finite critical point}
By imposing Eqs. (\ref{obvious}) on
(\ref{system1})-(\ref{system3}) and linearizing, we get
\begin{eqnarray}
\delta\dot\rho_x &=& - \frac{\Gamma}{2w}\, \, \delta\rho_T + \frac{16\pi\Gamma^4}{81(1 + w)^2w^2}\,\, \delta M \, , \\
\delta\dot\rho_T &=& \frac{\Gamma}{1 + w}\, \, \delta\rho_T + \frac{\Gamma w}{1 + w}\, \, \delta\rho_x \,  ,\\
\delta\dot M &=& 0 \, .
\end{eqnarray}
\\
The roots of the characteristic equation of this system are:
\\
\begin{equation}
\lambda_\pm = \frac{\Gamma}{2(1 + w)}\, [1 \pm \sqrt{- 1 - 2w}] \,
.
\end{equation}
\\
In view that $w < - 1$, both eigenvalues are real, one positive
and the other negative, i.e., a saddle point.

\subsection{The critical point at infinity}
A full analysis, in the original three-dimensional system, of the
critical point at infinity, is considerably hard since one has to
embed the system in a four-dimensional space of difficult
visualization.  We will consider, instead, the three
two-dimensional systems resulting from projecting the original one
upon three mutually orthogonal and complementary planes.

\subsubsection{The critical point at infinity in the plane
($\rho_{x}, \rho_T$)}

By setting  $M = 0$  the system (\ref{e1})-(\ref{e3}) gets
projected onto the plane ($\rho_{x}, \rho_T$), resulting
\\
\begin{eqnarray}
\dot x &=& - 3(1 + w)\sqrt{y}x - \Gamma x  = X(x,y) \, , \label{X1}\\
\dot y &=& - 3y^{3/2} - 3w \, \sqrt{y} \, x  = Y(x,y)\, ,
\label{Y1}
\end{eqnarray}
\\
where $x = \rho_x$ and $y = \rho_T$. This plane corresponds to the
situation that all black holes are formed with the same mass which
does not vary with time.

Let us introduce a new, ancillary, coordinate $z$, to deal with
points at infinity, and consider the unit sphere in the
three--dimensional space ($x,y,z$). Then, by defining new
coordinates, $u$ and $v$, by
\begin{equation}
x = \frac{u}{z} \, , \quad  \quad y = \frac{v}{z} \, , \quad
\quad u^2 + v^2 + z^2 = 1 \, ,
\end{equation}
we can write,
\begin{eqnarray}
X(x,y) &=& - 3(1 + w)\sqrt{y}x - \Gamma x \nonumber\\
&=& \frac{1}{z^{3/2}}\biggr[ - 3(1 + w)\sqrt{v}u - \Gamma u\,z^{1/2}\biggl] = \frac{1}{z^{3/2}}P(u,v,z) \, , \\
Y(x,y) &=& - 3y^{3/2} - 3w\, \sqrt{y} \, x \nonumber\\
&=& \frac{1}{z^{3/2}}\biggr[ - 3v^{3/2} - 3wv^{1/2}u\biggl]  =
\frac{1}{z^{3/2}}Q(u,v,z) \, .
\end{eqnarray}
\par
For the two--dimensional system  (\ref{X1})--(\ref{Y1}) one
follows
\\
\begin{equation}
- Y dx + X dy = 0 \, .
\end{equation}
Moreover,
\begin{equation}
dx = \frac{du}{z} - u\frac{dz}{z^2} \, , \quad \quad  {\rm and}
\quad \quad dy = \frac{dv}{z} - v\frac{dz}{z^2} \, .
\end{equation}

Hence
\begin{equation}
 Adu + Bdv + Cdz = 0 \, ,
\end{equation}
where
\begin{equation}
 A = - zQ \, , \quad  \quad B = zP \, , \quad  \quad C = uQ - Pv \, .
\end{equation}

Now, we  construct a new three--dimensional system
\\
\begin{eqnarray}
\dot{u} &=& Bz - Cv \, , \\
\dot{v} &=& Cu - Az \,  , \\
\dot{z}&=& Av - Bu \, .
\end{eqnarray}
\par

The region at infinity follows by setting $z = 0$. This implies
\\
\begin{equation}
uQ - vP = 0\, , \quad  \quad u^2 + v^2 = 1 \, ,
\end{equation}
i.e.,
\begin{equation}
u = 0\,  , \qquad  v= 1 \, , \qquad {\rm and} \qquad u = v =
\frac{\sqrt{2}}{2} \, .
\end{equation}
\\
These two are the  critical points at infinity.

Let us begin by considering  the first one, namely, $u = 0$, $v =
1$. To this end we perform the transformation,
\\
\begin{equation}
\xi = \frac{u}{v} \quad , \quad \eta = \frac{z}{v} \, ,
\end{equation}
and obtain the following system,
\\
\begin{eqnarray}
\dot\xi v + \xi\dot v &=& B\eta v - Cv \, ,\\
\dot v &=& C\xi v - A\eta v \, ,\\
\dot\eta v + \eta\dot v &=& Av - B\xi v \, .
\end{eqnarray}

Then, the two--dimensional system at infinity is,
\\
\begin{eqnarray}
\dot\xi &=& - C(1 + \xi^2) + B\eta  + A\eta\xi \, ,\\
\dot\eta &=& A(1 + \eta^2) - B\xi - C\xi\eta \, .
\end{eqnarray}
\\
Upon linearizing around the critical point at infinity ($\xi =
\eta = 0$, $v = 1$) we get,
\begin{eqnarray}
\dot\xi &=& - 3w\xi \, ,\\
\dot\eta &=& 3\eta \, .
\end{eqnarray}
\par
Altogether, this critical point at infinity is a saddle point for
$w > 0$ and a repeller for $ w \leq  0$.
\par
To study the second critical point, $u = v = \frac{\sqrt{2}}{2}$,
we perform a clockwise rotation so that the old  $u$ axis comes to
coincide with the new axis, $v'$. Proceeding  as before, we obtain
\begin{eqnarray}
\dot{\xi'} &=& 3 w \, \sqrt{\frac{\sqrt{2}}{2}}\; \xi' \quad , \\
\dot{\eta'} &=& 3(1 + w)\, \sqrt{\frac{\sqrt{2}}{2}}\; \eta' \, .
\end{eqnarray}
Clearly, this point is an attractor for $w < - 1$, a saddle for
$-1 < w <0 $, and a repeller for $0 < w $. Notice that the big rip
singularity corresponds to this point when $w < -1$.
\\
\begin{figure}[th]
\includegraphics[width=3.0in,angle=0,clip=true]{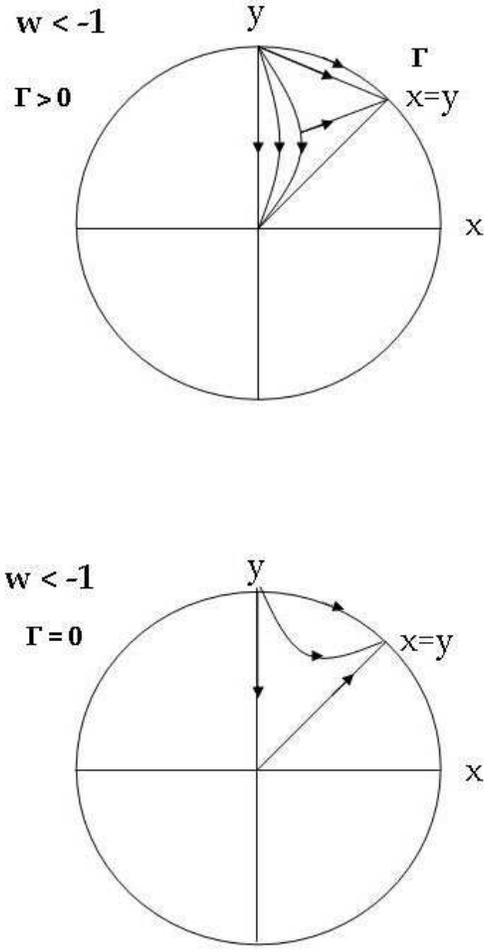}
\caption{Phase portraits of the system (\ref{X1})-(\ref{Y1}) when
the dark energy is of phantom type. The top panel corresponds to
the case of a non-vanishing rate of black hole formation. The
finite critical point (given by Eq. (\ref{obvious})), a saddle,
acts a divider: trajectories at its left, and some passing through
it, avoid the big rip; trajectories to its right, and some passing
through it, end up at the big rip -see text. The bottom panel
corresponds to the case of no black hole production. In this case,
all trajectories end up at the big rip.} \label{fig:phase1}
\end{figure}

The top panel of Fig. \ref{fig:phase1} displays the phase
portrait. All solutions start from $\rho_T = \rho_{bh} \rightarrow
\infty$ (i.e., the top point on the vertical axis, $x = 0$). Some
of them cannot avoid the big rip singularity (i.e., the common
point to the circle and the straight line $x = y$). Those
solutions that end up at the center of the circle (the Minkowski
state, $\rho_{x} = \rho_{bh} = 0$) evade the big rip. We remark,
by passing, that except for the particular case $\Gamma = 0$, no
solution can go from the origin to the infinity along the straight
line $\rho_x = \rho_T$.
\par
For vanishing $\Gamma$ (bottom panel of Fig. \ref{fig:phase1}) the
finite critical point collapses to the critical point at the
origin, which becomes a saddle point. This represents the usual
scenario of a system composed of pressureless and phantom fluids
in which is implicitly assumed that no black holes are produced.

\begin{figure}[th]
\includegraphics[width=2.50in,angle=0,clip=true]{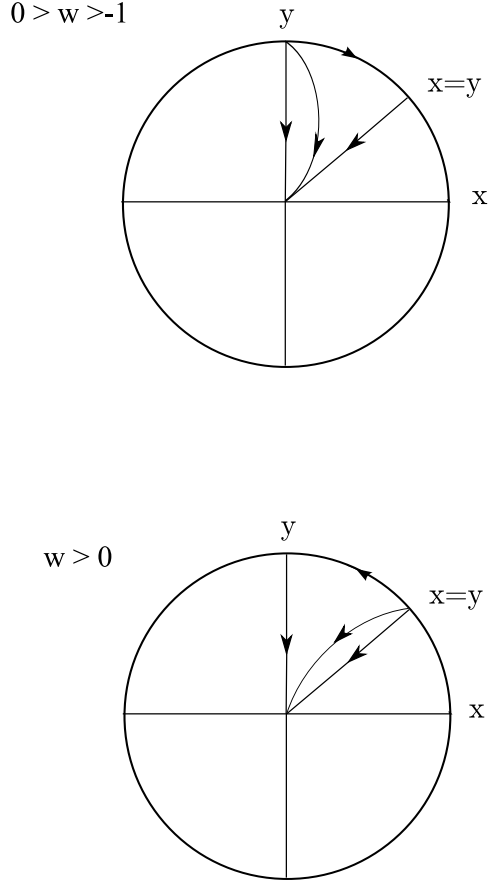}
\caption{Phase portrait of the system (\ref{X1})-(\ref{Y1}), with
$\Gamma = 0$, when the dark energy obeys the dominant energy
condition but fails to fulfill the strong energy condition (top
panel). For the sake of completeness, the bottom panel displays
the situation in which the fluid obeys both  energy conditions. In
both cases, because of the absence of phantom fluid, there is
neither black hole production nor big rip.} \label{fig:phase2}
\end{figure}

 The corresponding phase portraits when the dominant energy condition
 is satisfied are shown in the top ($0 > w > - 1$) and bottom
 ($ w > 0$) panels of Fig. \ref{fig:phase2}.
\\

\subsubsection{The critical point at infinity in the plane
$(\rho_{x}, M)$}
The plane $\rho_T = 0$ does not belong to the
physical region since it requires negative energy densities.
Anyway, for completeness let us analyze this case. After setting
$\rho_T = 0$, we get
\begin{eqnarray}
\dot x &=& - \Gamma x + (1 + w)\, y \, x^2 = X(x,y)\, ,\\
\dot y &=& (1 + w)\, y^2\, x = Y(x,y) \,
\end{eqnarray}
where $x = \rho_x$ and $y = 16\pi M$.
\par
Performing the transformations
\begin{equation}
u = zx \, , \quad  \quad v = zy \, , \quad  \quad u^2 + v^2 + z^2
= 1 \, ,
\end{equation}
it follows that,
\begin{eqnarray}
X(x,y) &=& \frac{1}{z^3}\biggr\{- \Gamma z^2u + (1 + w)u^2v\biggl\} = \frac{1}{z^3}P(u,v,z)\,  ,\\
Y(x,y) &=& \frac{1}{z^3}(1 + w)v^2u = \frac{1}{z^3}Q(u,v,z)\, .
\end{eqnarray}
As before, the relation
\begin{equation}
- Ydx + Xdy = 0
\end{equation}
implies
\begin{equation}
- zQ\, du +zP \, dv + (uQ - vP)\, dz = 0 \, .
\end{equation}

At infinity, $z = 0$, the relationship $uQ - vP = 0$ is
identically fulfilled. Hence, all points satisfying $u^2 + v^2 =
1$ are singular. This corresponds to a circle centered at the
origin. Since the latter is an attractor, all the trajectories
emanating from the infinity go to the origin to end there. So, the
infinity is a repeller.

\subsubsection{The plane $(\rho_{T}, M)$}
Upon setting $\rho_x = 0$ the original system reduces to
\\
\be
\dot\rho_T = - 3\, \rho_{T}^{3/2} \, , \quad  \quad \dot{M}=0
\, .
\ee
\\
which admits the simple solution, $\rho_T \propto t^{-2}$. Hence,
the solution comes  from the infinity to the origin along the axis
$\rho_{x} = 0$.

\section{Discussion}
In this paper, we briefly raised the  point that black holes may
be produced in phantom dominated universes  by three different
mechanisms:   $(i)$ via energy accumulation in any given spatial
three-volume, $(ii)$ gravitational collapse of huge thermal
fluctuations, and $(iii)$ quantum tunnelling of very dense and hot
phantom fluid \cite{quantumtunnelling}. In this regard, phantom
fluids might be viewed as ``black holes factories". However, while
these processes look rather plausible the corresponding
calculations are pending.

Nevertheless, after accepting that black holes may be produced by
any of the sketched mechanisms, we studied the dynamical system
associated to this scenario. Our main finding is that because of
the existence of a critical saddle point in the finite region of
the plane $(\rho_{x},\rho_T)$ -given by Eqs. (\ref{obvious})- the
big rip singularity (a generic feature of phantom-dominated
universes) is no longer unavoidable. This critical point lies in
the region of positive densities as $ w < 0$. Moreover, since
$\rho_T \geq \rho_x$  and $\rho_T/\rho_x = - w $, this point is
located in the physical region ($\rho_T \geq \rho_x > 0$) when $w
< - 1$, that is, for  phantom equations of state.
\par
In the said plane $(\rho_{x},\rho_T)$, there is an attractor at
the origin, also the finite critical point described above (a
saddle), and two critical points at infinity: one situated at the
axis $\rho_T$ (a repeller), and another at the straight line
$\rho_T = \rho_x$ (an attractor). The finite critical point,
connected with $\Gamma > 0$, is a divider between the solutions
that go from $\rho_x \rightarrow \infty$ to the origin and from
$\rho_x \rightarrow \infty$ to the big rip. Those solutions that
pass through this critical point, depending on the initial
conditions, can either go to the origin (i.e.,  implying that the
big rip is avoided), or to the critical point at infinity with
$\rho_x = \rho_T$ (i.e., big rip) -see top panel of Fig.
\ref{fig:phase1}.
\par
We are now in conditions to fix more precisely the possible
scenario described above. Let us consider the relations for the
finite critical point, Eqs. (\ref{obvious}). Bearing in mind that
the Universe is nearly spatially flat ($\Omega_{T} =1$), and
conceding that this critical point may lie not very far from the
present expansion era (the Universe began accelerating recently),
we find $\Gamma = - 3\sqrt{3}\, (1 +w)H_0$, and $\Omega_{x0} =
-w^{-1}$. The Wilkinson microwave anisotropy probe (WMAP)
\cite{wmap} data are consistent with $w \simeq - 1.1$. Hence, for
the big rip to be attained the present dark energy density
parameter must fulfill $\Omega_{x0}  \gsim \, 0.9$. Since WMAP
indicates $\Omega_{x0} \simeq 0.7$, our Universe may well avoid
the big rip singularity (modulo $w$ is really a constant). This
also implies $\Gamma \sim 0.5 \,H_0$.
\par
Our system is a three--dimensional one and obviously there are
other dimensions. In the plane $(\rho_{x},M)$ all points at
infinity are singular; projecting this onto the Poincar\'{e}
sphere we obtain, in that plane, a singular circle around the
origin. But, since the origin is an attractor, all the points in
this circle are repellers.
\par
As said above, some solutions can evade the big rip. Obviously,
the latter become unavoidable if $\Gamma = 0$ (no black hole
production), since in this case the critical point at the finite
region coincides with the origin.

There are also other situations in which the big rip can be
avoided. For instance, when phantom dark energy corresponds to the
generalized Chaplygin gas proposed in \cite{pedro}, or when
wormholes intervene \cite{jimenez}, or when the curvature scalar
gets very large and quantum effects become dominant
\cite{elizalde}. Nevertheless, to the best of our knowledge, the
present scenario was never considered in the literature.

By contrast, as noted by Barrow \cite{jdb}, there are situations
in which finite-time future singularities can arise even if the
fluid filling the Universe obeys $\rho > 0$ and $\rho + 3p >0$,
i.e., under very mild conditions. We do not consider them here.

Admittedly, it can be argued that in view of the various
simplifying assumptions, our treatment is not much realistic. In
the first place, the rate $\Gamma$ is not expected to be a
constant, it will likely vary with expansion and depend on
quantities like $w$ and $M$. Secondly, we have implicitly
considered that all black holes are formed simultaneously with the
same mass -a flat spectrum. It would be more natural to assume the
number of black holes produced varies with mass and time. Further,
as noted earlier, black hole spontaneous radiance should be
included. Clearly, these features ought to be incorporated in
future, more realistic, treatments. Nonetheless, we believe this
small, first, step may lead the way to more ambitious
undertakings.


\acknowledgments{One of us, DP, wishes to thank to the Department
of Physics of the Universidade Federal do Esp\'{\i}rito Santo,
where this work was started, for warm hospitality, and the CNPq
(Brazil) for financial support. This research was  partially
supported by the Spanish Ministry of Education and Science under
Grant FIS2006-12296-C02-01, and the ``Direcci\'{o} General de
Recerca de Catalunya" under Grant 2005 SGR 00087. Likewise, J.C.F.
thanks FAPES(Brazil), CNPq (Brazil) and the French-Brazilian scientific
cooperation program CAPES-COFECUB for partial financial support.}

\end{document}